# Trajectory and orbit of the unique carbonaceous meteorite Flensburg

Jiří BOROVIČKA[1*], Felix BETTONVIL[2], Gerd BAUMGARTEN[3], Jörg STRUNK[4], Mike HANKEY[5], Pavel SPURNÝ[1], and Dieter HEINLEIN[6]

[1] Astronomical Institute of the Czech Academy of Sciences, Fričova 298, CZ-25165 Ondřejov, Czech Republic
[2] Leiden Observatory, Leiden University, Niels Bohrweg 2, 2333 CA Leiden, The Netherlands
[3] Leibniz-Institute of Atmospheric Physics at Rostock University, Schlossstraße 6, D-18225 Kühlungsborn, Germany
[4] European Fireball Network and Arbeitskreis Meteore, Herford, Germany
[5] American Meteor Society LTD, USA
[6] German Fireball Network, Lilienstraße 3, D-86156 Augsburg, Germany

*Corresponding author. E-mail: jiri.borovicka@asu.cas.cz

**Abstract** – The C1-ungrouped carbonaceous chondrite Flensburg fell in Germany on September 12, 2019, in the daytime. We determined the atmospheric trajectory, velocity, and heliocentric orbit using one dedicated AllSky6 meteor camera and three casual video records of the bolide. It was found that the meteorite originated in the vicinity of the 5:2 resonance with Jupiter at heliocentric distance of 2.82 AU. When combined with the bolide energy reported by the U.S. Government sensors (USGS), the pre-atmospheric diameter of the meteoroid was estimated to 2 – 3 meters and the mass to 10,000 – 20,000 kg. The meteoroid fragmented heavily in the atmosphere at heights of 46 – 37 km, under dynamic pressures of 0.7 – 2 MPa. The recovery of just one meteorite suggests that only a very small part of the original mass reached the ground. The bolide velocity vector was compared with that reported by the USGS. There is good agreement in the radiant but the velocity value has been underestimated by the USGS by almost 1 km s$^{-1}$.

## INTRODUCTION

The Flensburg meteorite fell in northern Germany, close to the Danish border, on September 12, 2019, 12:50 UT (14:50 local daylight saving time). The fall was accompanied by a very bright bolide (a superbolide) easily visible in broad daylight. The International Meteor Organization received 584 reports of visual sightings of the bolide through its online form (https://fireballs.imo.net/). Most of the reports were from the Netherlands, where the weather was best at the time of the event. The most distant sighting, more than 600 km from the bolide, was reported from England.

The bolide was also detected by the space borne U.S. Government sensors (USGS, https://cneos.jpl.nasa.gov/fireballs/, see also Brown et al. 2016). The sensors provided the exact time of the bolide (12:49:48 UT), approximate geographic location (54.5° N, 9.5° E) and height (42 km) of the point of maximum brightness, bolide velocity (18.5 km s$^{-1}$) and total radiated energy (1.69 × 10$^{11}$ J). The total impact energy was deduced to be 0.48 kt TNT, i.e. 2.0 × 10$^{12}$ J. USGS report only about a dozen of bolides with similar or larger energies per year globally. The event was therefore quite significant on regional scale. The provided velocity components indicated that the bolide moved nearly from the south to the north (heading azimuth 9°) on a trajectory with low slope (24° to the horizontal). Note, however, that while the USGS reported energies can be considered as reliable, the velocities and trajectories were found to be in severe error for some bolides with independent data available (Borovička et al. 2017, Devillepoix et al. 2019).

A single meteorite fragment was found by chance already the next day, September 13, by Mr. Erik Due-Hansen on the lawn of his yard in Flensburg. The meteorite has been described and analyzed by Bischoff et al. (2020). The total mass was 24.5 g with the bulk density of only 1984 kg m$^{-3}$. The meteorite has been classified as C1-ungrouped carbonaceous chondrite. It therefore represents a unique material not found in other known meteorites. No further meteorite fragment from the Flensburg fall was found as of the time of writing (September 2020).

The purpose of this paper is to set the Flensburg meteorite into a geological context within the solar system, i.e. to determine its pre-fall heliocentric orbit and possible source region. At the same time, the atmospheric trajectory of the bolide will be studied with the aim to obtain information about the physical properties of the original meteoroid. Since we cannot rely on the USGS data, we will use four ground-based video records of the bolide. One video was obtained by us and three others were found on the internet and carefully calibrated. Similar work has been done for three other instrumentally observed carbonaceous meteorite falls: C2-ungrouped Tagish Lake (Brown et al. 2000, Hildebrand et al. 2006) and two CM2 meteorites, Sutter's Mill (Jenniskens et al 2012) and Maribo (Borovička et al. 2019).

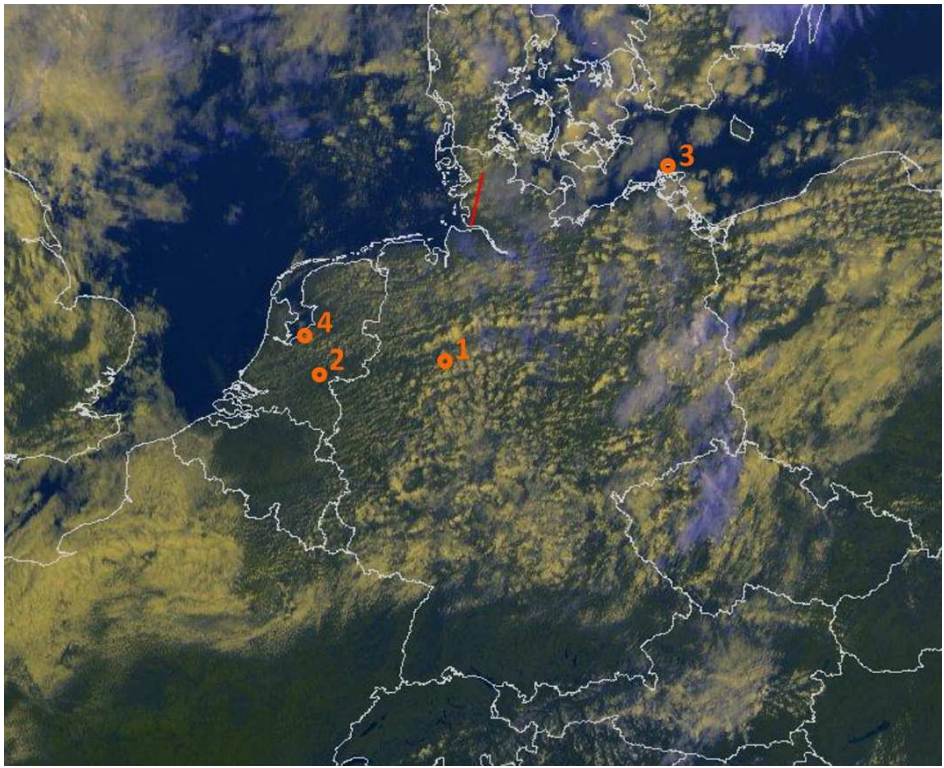

Fig.1. Image of Germany and surrounding areas taken by the meteorological geostationary satellite Meteosat on 12 September 2019, 12:45 UT (courtesy EUMETSAT and Czech Hydrometeorological Institute). Ground projection of the bolide trajectory (red line) and locations of the four videos used for the trajectory determination are indicated.

# THE VIDEO RECORDS

At the time of the bolide, three meteor observing camera systems AllSky6 were in operation in Germany. The AllSky6 system (Hankey et al. 2020) consists of six video cameras at each site, which run 24 hours a day. As is illustrated on an image from the meteorological satellite Meteosat in Fig. 1, the early afternoon of September 12 was partly or mostly cloudy over the majority of Germany and the surrounding countries. Only the camera in Herford, operated by one of us (JS), recorded part of the bolide, about one second long, in a gap between clouds (Fig. 2). Since the camera is stationary, it was easy to calibrate the video astrometrically. Stars were measured on four saved video sequences (showing other meteors) from the night of September 10/11 and three sequences from September 13/14. All data fit well together, confirming that the camera did not move in between. In total, 491 star positions were used. Nine plate constants of the axially symmetric projection described in Borovička (2014) were determined. The plate coordinates $x,y$ could then be transformed into azimuths and elevations $A,h$.

Several casual video records of the bolide were published on the Internet. From these, we selected three videos which were suitable for on-site stellar calibration and showed the bolide from different angles. All three videos were recorded by dashboard cameras in moving cars. The car motion complicated the calibration, nevertheless, we were able to extract the bolide coordinates well, especially from two videos. We also tried to calibrate another video taken by a stationary security camera in Marknesse, the Netherlands, but the low quality of the camera and lack of sufficient terrestrial calibration objects prevented us to obtain useful data from that record. The camera scale was different for x- and y-axis and we were unable to obtain the correct ratio.

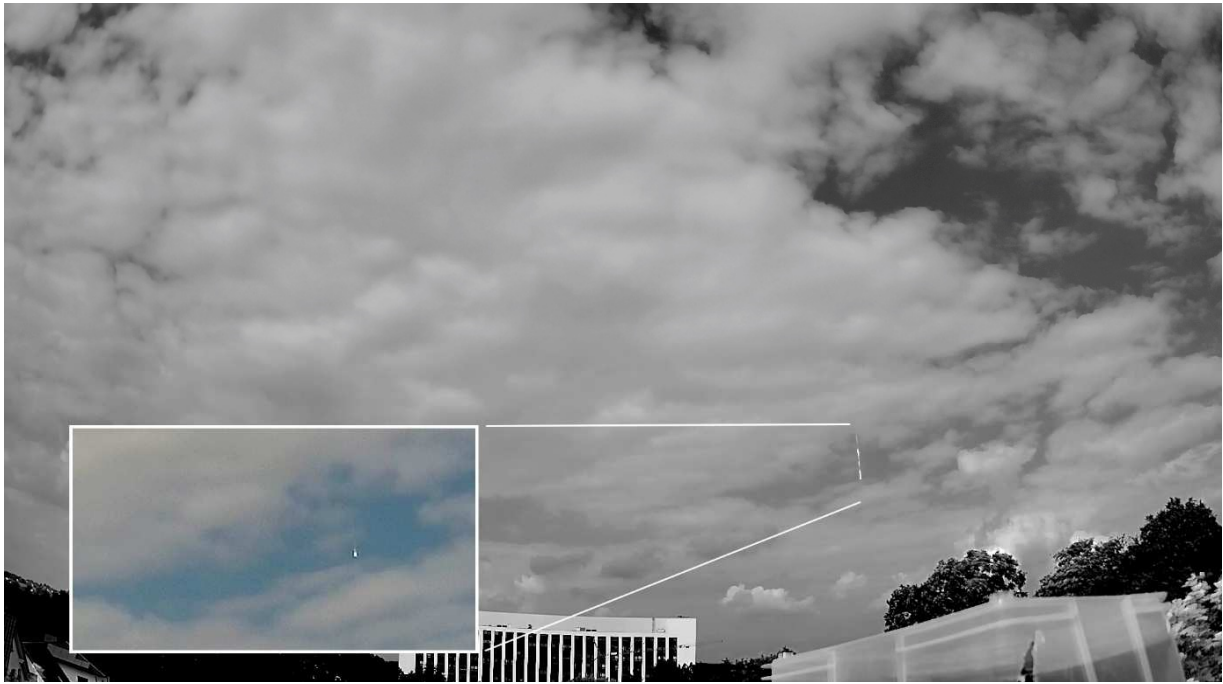

Fig. 2. Co-added video frames from the Herford video, converted to greyscale, showing the bolide path in a gap between clouds. The color inset shows an enlargement from a single frame. Author of the video: Jörg Strunk.

A very useful video was taken on the Energieweg in Nijmegen, the Netherlands. Some images from the video are shown in Fig. 3. The bolide appeared as a small spot near the center of the frame. It brightened slowly before flaring up suddenly. After the flare, the main fragment continued down, leaving a bright trail behind it, visible on several subsequent frames. The whole bolide, lasting for 4.5 seconds, passed on a clear part of the sky, although a few frames, where the bolide crossed high-voltage aerial power cables, were not measurable. The video frame shown in Fig. 3 was calibrated astrometrically using the method described in Borovička (2014). The night time starry images were taken from nearly the same spot by one of us (FB) on December 18, 2019. The left third of the image was to be covered because of too bright street lights. Still, sufficient number of stars (38) could be measured. At the same time, 23 terrestrial objects, well visible on both the original video and the calibration image, were measured. Their distances from the calibration camera were measured on Google Earth and ranged from 30 meters for nearby objects to almost 2 kilometers for a distant bridge construction. The difference between the position of the calibration camera and the actual position of the original video camera was found by minimizing the residua of the calibration. The difference was about 1.3 meters, nearly in the direction of car motion. This terrestrial coordinate correction was taken into account in the final astrometric calibration (see Borovička 2014 for details of the calibration procedure).

Since the car did not move straight all the time, the aiming point of the camera changed. The changes were determined by measuring the distant bridge on each video frame. The plate constants $A_0$, $h_0$ (azimuth and elevation of the aiming point) were adjusted accordingly for each frame. Other plate constants were kept fixed. The distance covered by the car during the bolide flight was ignored since it was small in comparison with the scatter of the measurements at the bolide distance (in fact only the projection of the car motion perpendicular to the line of sight may play a role and that was quite small because the car moved toward the bolide).

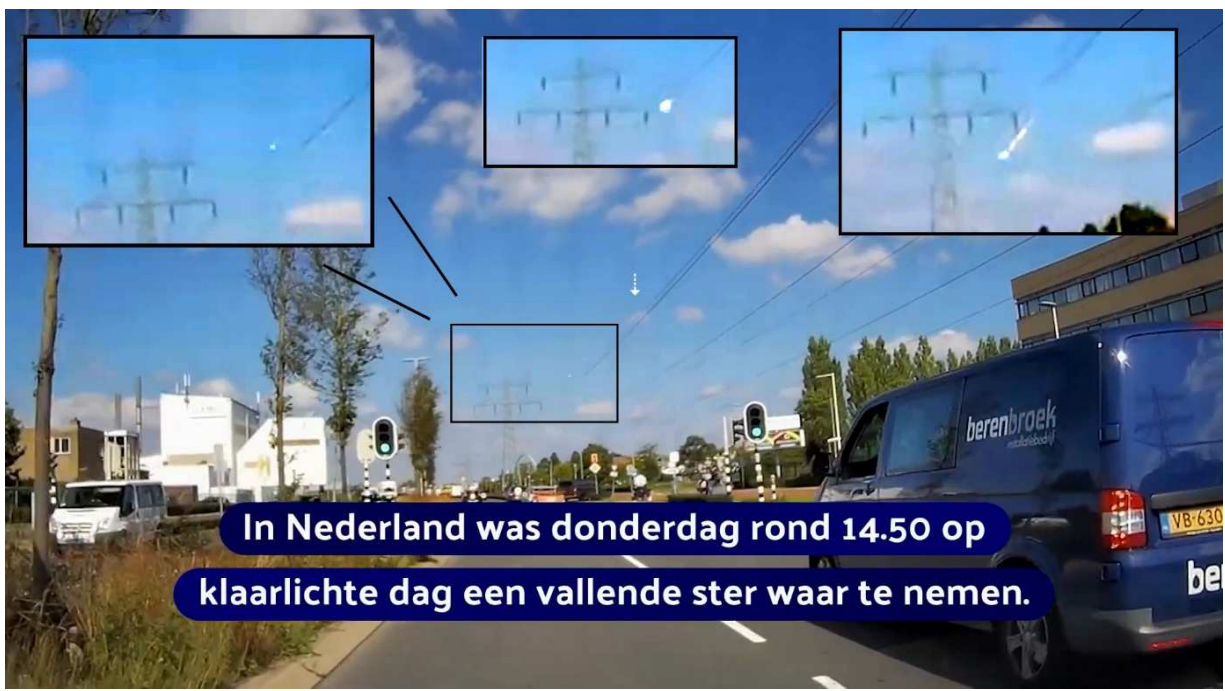

Fig. 3. Single frame from the Nijmegen video. This frame was used for terrestrial calibration. The bolide is nearly in the middle, better visible in the inset at upper left, where part of the image was enlarged and contrast was enhanced. The white arrow above the center of the frame indicates where the bolide started to be visible 2.5 seconds earlier. The other two insets show the bolide in the maximum light and toward the end. Author of the video: Gerard Kemna. The video was downloaded from the site of the Dutch news media NU.nl (https://www.nu.nl/binnenland/5991876/heldere-vallende-ster-op-klaarlichte-dag-waargenomen-boven-nederland.html)

Another useful video was taken at a road near the northern tip of the German island Rügen. Some frames of this video are presented in Fig. 4. Also here the bolide was located near the center of the field of view, which is convenient because image distortion is negligible here. At the beginning, the bolide emerged from behind the clouds and the bolide end was also hidden by clouds. Nevertheless, a significant part of the bolide, about three seconds long, was observed, though also that part was interrupted three times by clouds. The calibration image was taken by one of us (GB) on March 11, 2020. In total, 111 stars and 24 terrestrial objects in distances from 30 m to 2.7 km were measured. No correction of camera position was necessary. On the other hand, since the tilt of the video camera changed during the car motion (as the driver was passing a cyclist), three plate constants, $A_0$, $h_0$, and $\varphi$, were to be adjusted for each frame using two points on the distant horizon.

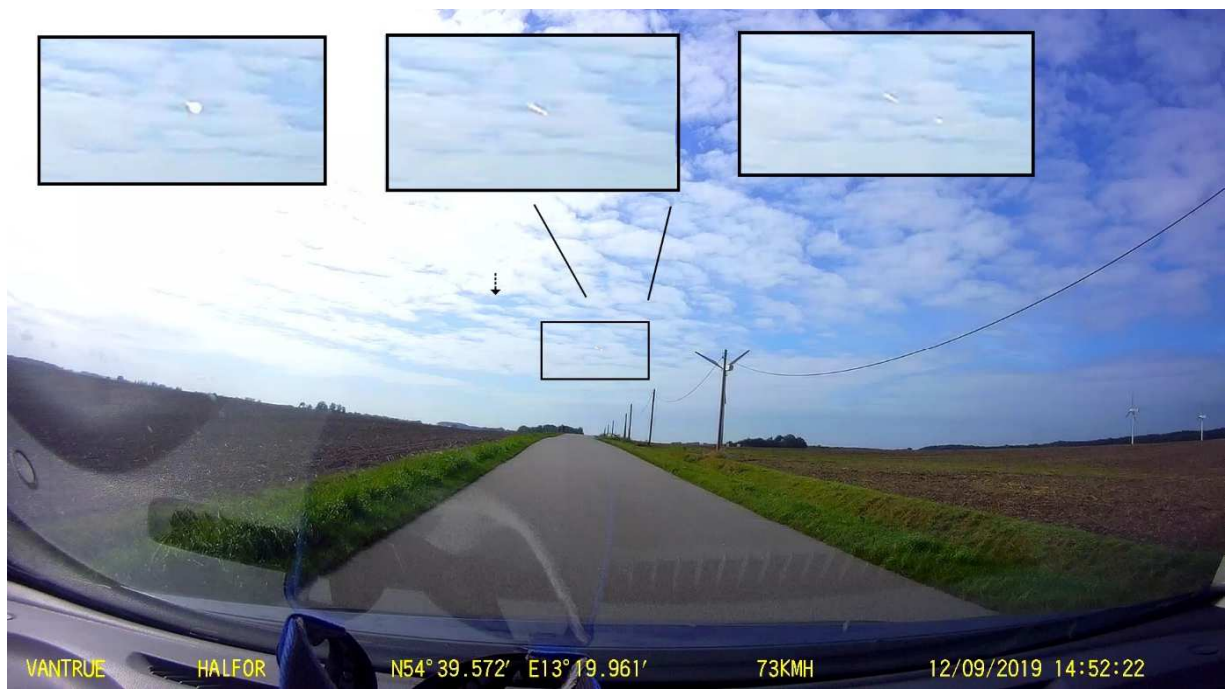

Fig. 4. Single frame from the Rügen video. This frame was used for terrestrial calibration. The bolide is nearly in the middle, better visible in the middle inset, where part of the image was enlarged. The black arrow indicates where the bolide started to be visible 3 seconds earlier. The left inset shows the bolide at maximum brightness, which occurred just 6 frames earlier. The right inset shows one of the last frames, where the bolide is visible. Author of the video: Holger Scheele. The video was downloaded from https://www.youtube.com/watch?v=W0ZFqBDdTJk.

Finally, a video taken in Almere, the Netherlands, was calibrated (Fig. 5). The bolide started to be visible on a clear part of the sky but disappeared behind clouds one second later. After another two seconds, the bolide reappeared shortly (on three frames) in a small gap between clouds. Since the car turned on roundabout in between, both parts of the video were calibrated independently using separate calibration images taken by one of us (FB) on July 11, 2020. There were 76/78 stars and 21/18 terrestrial objects in distances from 10 m to 0.5 km. Corrections of camera positions were applied. Because of lower quality of this video and less convenient terrestrial calibration objects, the calibration of this video was less robust than was the case of previous two casual videos. A range of possible solutions, varying in the values of camera position correction and by including/excluding some of the calibration objects, was obtained. The selected solution is fully consistent with the bolide trajectory and velocity determined from other three videos. Other formally allowable solutions resulted in deviations of lines of sight from the bolide trajectory up to about ± 1 km. The bolide trajectory presented in the next section was therefore primarily based on the videos from Herford, Nijmegen, and Rügen. The Almere video demonstrated that there is no conflict with that solution.

The parameters of all four videos including geographical coordinates are given in Table 1. The video locations relative to the bolide path are shown in Fig. 1.

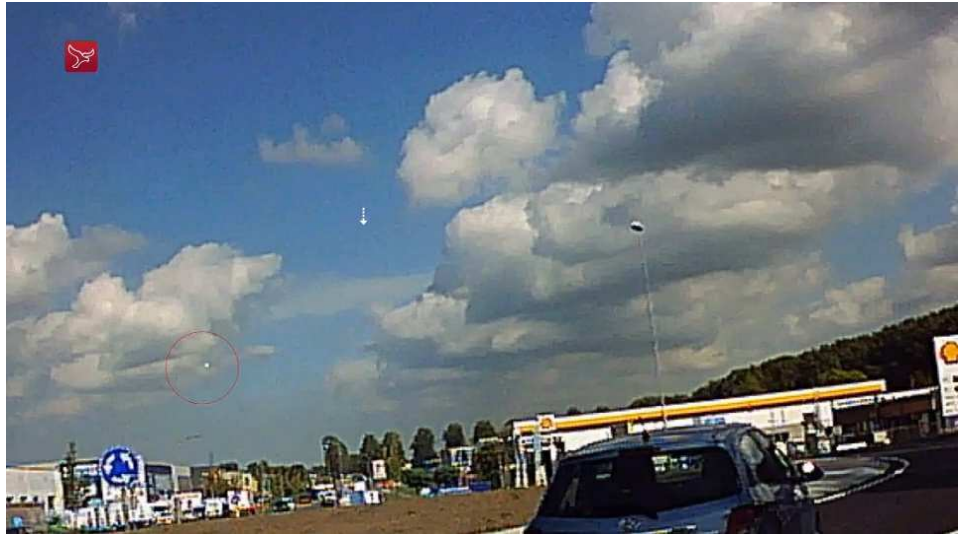

Fig. 5. Single frame from the Almere video. This frame was used for terrestrial calibration of the second part of the video, when the bolide (in red circle) shortly reappeared between clouds. The white arrow indicates where the bolide started to be visible 3 seconds earlier. Author of the video: Leon Pepping. The video was downloaded from the site of the Dutch news media omroepflevoland.nl (also available at https://www.youtube.com/watch?v=Qp2_BZju7EU).

Table 1. Video records used for trajectory determination

| No. | Source | Site | Longitude | Latitude | Altitude | Resolution | FOV | FPS | NOF |
|---|---|---|---|---|---|---|---|---|---|
| 1 | Meteor camera | Herford | 8.70178 | 52.12364 | 121 m | 1920×1080 | 87°×47° | 25 | 23 |
| 2 | Moving car | Nijmegen | 5.82536 | 51.84390 | 8 m | 1920×1080 | 68°×40° | 25 | 106 |
| 3 | Moving car | Rügen | 13.33233 | 54.65956 | 17 m | 1920×1080 | 95°×58° | 30 | 56 |
| 4 | Moving car | Almere | 5.26111 | 52.40027 | 0 | 960×540 | 46°×25° | 25 | 24 |

The last three columns contain the estimated field of view (FOV), the number of frames per second (FPS) and the number of frames with visible bolide (NOF)

## BOLIDE TRAJECTORY AND VELOCITY

The bolide trajectory was computed by the least squares method of Borovička (1990). Video calibrations provided the source data: bolide azimuths and elevations for each video frame. The source data can be found in the Supplementary file Flensburg.xlsx. In Herford, the bolide was visible at elevations 17° – 14° above the horizon. In all other videos, the first measurements were at elevations between 11° and 12°. The end of the bolide was visible only from Nijmegen, 3.5° above the horizon.

The trajectory was first assumed to be straight. Corrections for curvature due to gravity were applied at the end, when the linear trajectory and velocity were known. The least squares method finds the linear trajectory by minimizing the sum of squares of the distances (in space) between the trajectory and the lines of sight. Since the videos and their calibrations were of varying quality, weights were set differently for each video. Herford got the weight 100, Nijmegen and Rügen 1, and Almere 0.1. Nevertheless, the results did not differ much for equal weights: by less than 0.1 degree for the radiant and by about 100 m for the position of the end point. Good consistency of all measurements is demonstrated in Fig. 6, where the deviations of lines of sight from the trajectory are plotted. There is no significant systematic trend of any video. The points are randomly mixed. For Herford, most points lie within 50 meters of the trajectory. The scatter of the other videos is about ± 300 meters. This larger scatter is mostly due to the fact that the cameras were located in moving cars.

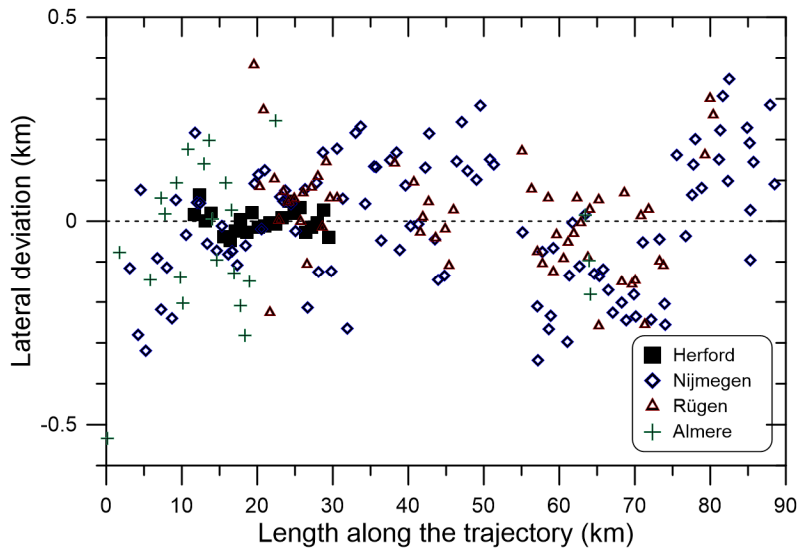

Fig. 6. Deviations of individual lines of sight from four video sites from the bolide trajectory. The deviation is positive if the line of sight passes above the trajectory or, in the case of Herford, which lay almost in the fall plane, east of it. Note the different scales on the x- and y-axes.

The observed trajectory was 90 km long. The bolide was first detected at a height of 71.8 km from Almere. From Nijmegen, the first detection was at a height of 70.5 km. The Herford camera covered a stretch 18 km long between heights 67.0 and 59.6 km. The last observation was from Nijmegen at a height of 35.3 km. The average slope of the trajectory to the horizontal was 24.4° (the slope changes along the trajectory due to Earth's curvature).

After the trajectory was computed, individual measurements were projected onto it. The dependency of the length along the trajectory (i.e. the distance travelled from the bolide beginning) as a function of time was obtained this way. The relative time was computed for each video from the known frame rate and considering that the cameras had a rolling shutter (i.e. the frames were not read out at once but line by line). These relative times are provided in the Supplementary file Flensburg.xlsx. The time offsets between cameras were then determined using the bolide data. The following time offsets relative to the Herford camera were found: Nijmegen -0.4549 s, Rügen 0.4121 s, Almere -0.5556 s. Using the time of bolide maximum reported by the USGS, the time zero at Herford was found to correspond to 12:49:45.2 UT.

From the Herford data alone, the average velocity of the bolide in the covered stretch was 19.422 ± 0.020 km s$^{-1}$. The data from the other videos are fully consistent with this value but show a larger scatter. In Fig. 7 the difference between the observed length and the expected length at the given time is plotted. Almere and Nijmegen show the largest scatter, Rügen shows smaller deviations. The scatter grows near the end of the trajectory, where the bolide was fragmented and more difficult to measure. Again, there is no systematic trend among the videos. Note that the velocity is not expected to remain constant along the whole trajectory. Atmospheric drag causes deceleration even for bolides of this size. The lag expected from bolide modeling is indicated by the dotted line in Fig. 7.

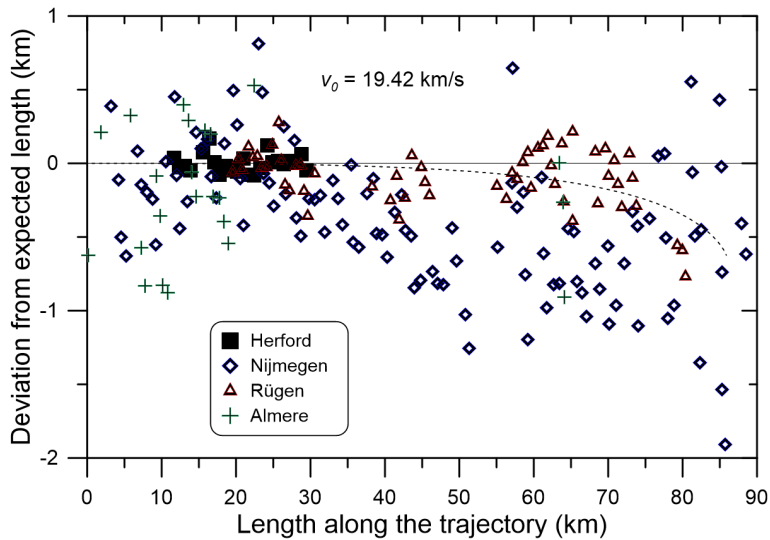

Fig.7. Deviations of the measured lengths along the trajectory from the length expected for the given time and a constant bolide velocity of 19.42 km s$^{-1}$. The dotted line shows the expected deceleration.

Since the flight was rather horizontal, the Earth's gravity must have caused some bending of the trajectory. It can be calculated that after the 4.55 seconds covered by the observations, gravity will displace the meteoroid about 90 meters in lateral direction. This is less than the scatter of the data (see Fig. 6), so the trajectory curvature could not be directly measured. Nevertheless, the expected change of the radiant was taken into account when computing the heliocentric orbit.

Trajectory, radiant, and velocity data are provided in Table 2. The beginning and end points are the points where the bolide started and ceased to be visible on the videos, respectively. Should the bolide appear at night, it would certainly have started to be visible at larger height and could be probably followed to somewhat lower height. The errors of the heights quoted in Table 2 express the uncertainty in the position and orientation of the trajectory, not the uncertainty of visibility of start/end point. The point of maximum brightness was not well defined since the bolide exhibited similar brightness at the heights of 43 – 40 km. The entry speed takes into account the expected slight atmospheric deceleration before the bolide appeared on the Herford video.

Table 2. Trajectory, apparent radiant and velocity, including the comparison with the USGS data

|  | Longitude (deg E) | Latitude (deg N) | Height (km) | |
|---|---|---|---|---|
| Beginning point | 9.0322 ± 0.0006 | 53.887 ± 0.002 | 71.84 ± 0.08 | |
| Maximum brightness | 9.180 | 54.492 | 40.7 | |
| *USGS* | *9.5* | *54.5* | *42* | |
| End point | 9.2060 ± 0.0016 | 54.598 ± 0.002 | 35.3 ± 0.08 | |
|  | Azimuth | Zenith distance | Right ascension | Declination |
| Apparent radiant* | 188.1° ± 0.1° | 65.3° ± 0.1° | 185.00° ± 0.10° | -11.01° ± 0.12° |
| *USGS* | *188.8°* | *66.4°* | *184.7°* | *-11.4°* |
| Entry velocity* | 19.43 ± 0.05 km/s | | | |
| *USGS* | *18.5 km/s* | | | |

*Apparent radiant is given at the beginning of the trajectory. Entry velocity is given at the top of the atmosphere. USGS data are valid for the maximum brightness point. Azimuth is counted from the north clockwise. Equatorial coordinates are for the equinox J2000.0

For comparison, data from the USGS are given in Table 2 as well. We note that the USGS position is found about 20 km to the east. The radiant agrees within one degree. Note that radiant azimuth and zenith distance change along the trajectory due to Earth's curvature. Right ascension and declination change only slightly due to the curvature of the trajectory. The biggest discrepancy to the USGS data is the bolide velocity, which was reported to be almost 1 km s$^{-1}$ lower than precisely deduced in this work. The difference is only partly explained by atmospheric deceleration between the beginning and the maximum brightness point. The measured average velocity between the heights of 45 – 41 km (from the Nijmegen video) was 19.3 ± 0.5 km s$^{-1}$ and the modeled velocity at 42 km was 19.1 km s$^{-1}$, i.e. still 0.6 km s$^{-1}$ larger than the USGS value.

Figure 8 shows the ground projection of the trajectory on the map. The bolide appeared above the Elbe river estuary and then continued nearly northwards, above land.

The geocentric radiant and heliocentric orbit were computed from the apparent radiant and entry velocity by the analytical method of Ceplecha (1987). The results are given in Table 3. The orbit is compared graphically with the orbits of three other carbonaceous chondrites with known orbit in Fig. 9. The orbit of Flensburg, whose fall was captured with one dedicated meteor camera, is probably the most precise among them while the orbit of the Tagish Lake, which was not directly imaged from the ground, is the least reliable. All orbits lie inside the orbit of Jupiter. The inclinations are small in all cases. The inclination of Flensburg (6.8°) is the largest of all four. Also the aphelion distance of Flensburg (4.8 AU) is the largest one but is comparable with Maribo (4.4 AU) and Sutter's Mill (4.7 AU). These two CM2 chondrites have lower perihelia, higher eccentricities, and lower semimajor axes than Flensburg. The most significant aspect of the Flensburg orbit is that it was in 5:2 resonance with Jupiter. This powerful resonance causes one of the Kirkwood gaps in the main asteroid belt, where all orbits become chaotic (e.g. Moons 1996). Although the Tisserand parameter of 2.89 with respect to Jupiter may suggest a cometary orbit, the fact that it was in the 5:2 resonance allows us to classify the Flensburg orbit as asteroidal (Tancredi 2014).

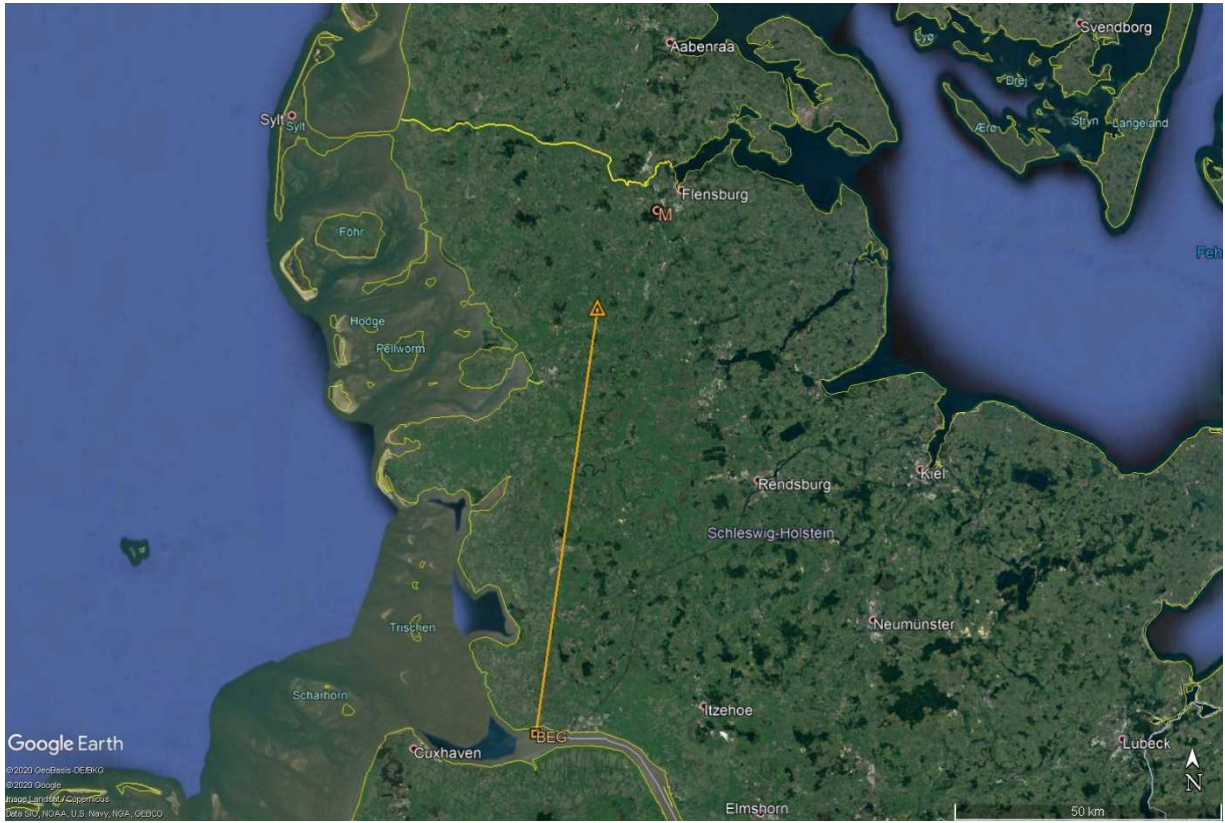

Fig. 8. Ground projection of the bolide trajectory and the position of the meteorite. The source of the background image is Google Earth.

Table 3. Geocentric radiant and heliocentric orbit (J2000.0)

| | | |
|---|---|---|
| Geocentric right ascension of radiant | $\alpha_g$ | 183.46° ± 0.11° |
| Geocentric declination of radiant | $\delta_g$ | −18.18° ± 0.14° |
| Geocentric entry speed (km/s) | $v_g$ | 15.97 ± 0.06 |
| Semimajor axis (AU) | $a$ | 2.82 ± 0.03 |
| Eccentricity | $e$ | 0.701 ± 0.003 |
| Perihelion distance (AU) | $q$ | 0.843 ± 0.001 |
| Argument of perihelion | $\omega$ | 307.25° ± 0.16° |
| Longitude of ascending node | $\Omega$ | 349.207° ± 0.001° |
| Inclination | $i$ | 6.82° ± 0.06° |
| Aphelion distance (AU) | $Q$ | 4.80 ± 0.06 |
| Perihelion date | | 2019-08-07.6 ± 0.1 d |
| Tisserand parameter | $T$ | 2.89 ± 0.02 |
| Orbital period | $P$ | 4.74 ± 0.08 yr |
| | $P_{jup}/P$ | 2.502 ± 0.007 |

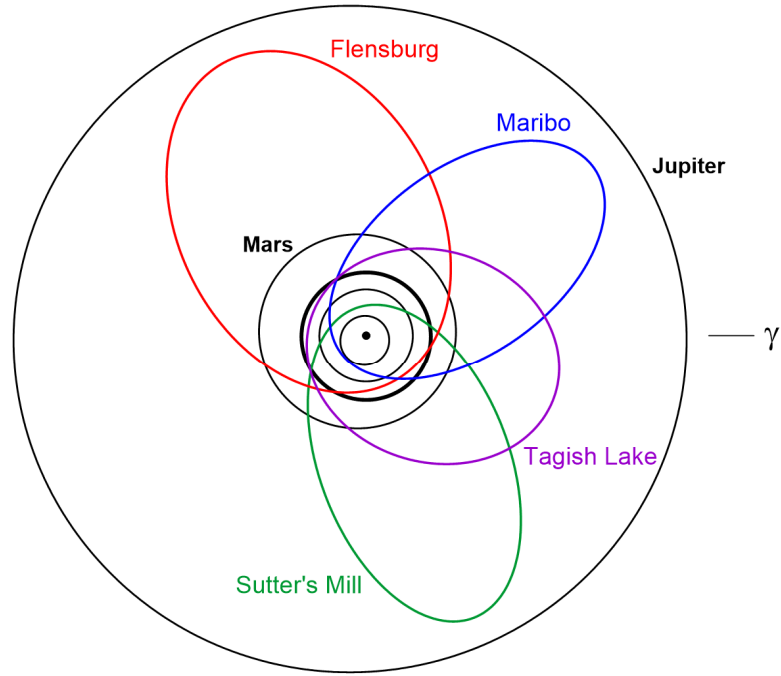

Fig. 9. Known orbits of carbonaceous chondrites in projection onto ecliptic plane: Tagish Lake (C2 ungrouped, fell in 2000, orbital data from Hildebrand et al. 2006), Maribo (CM2, 2009, Borovička et al. 2019), Sutter's Mill (CM2, 2012, Jenniskens et al. 2012), and Flensburg (C1 ungrouped, 2019, this work).

## LIGHT CURVE AND FRAGMENTATION

Since the fall occurred during the daytime, a detailed radiometric light curve, such as the case for the Maribo fall (Borovička et al. 2019), is not available. To get at least an approximate light curve, we used the Nijmegen video. Assuming that the camera has a gamma factor of 0.45, the bolide signal on each video frame was measured. An ad hoc correction for signal saturation was applied for the period of large brightness. Finally, the light curve was scaled to get the total radiated energy of $1.69 \times 10^{11}$ J as reported by the USGS. The resulting light curve is presented in Fig. 10. It must be kept in mind that the way to obtain it was not rigorous. Because the values of absolute magnitude and the exact shape of the light curve could not be determined independently, their reliability is limited. Nevertheless, the main characteristics are obvious. The brightness was increasing steadily from the beginning down to the height of 45 km, where the bolide exploded, i.e. the brightness increased abruptly. The increase was followed be a broad maximum at heights 45 – 40 km. After that, the brightness decreased, possibly with some minor flares on the descending branch.

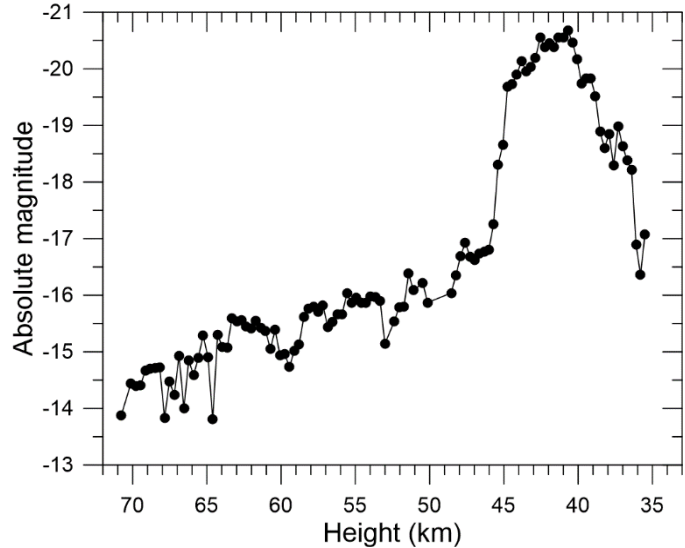

Fig. 10. Approximate light curve of the Flensburg bolide based on the Nijmegen video.

The light curve is very different from that of Maribo, which exhibited a steep increase at the beginning and many relatively short flares in the second half of the trajectory (Borovička et al. 2019). Maribo fragmented many times, which was interpreted as an evidence for inhomogeneous nature of the meteoroid. The meteoroid contained parts with various strengths, which disintegrated at various dynamic pressures. The initial disruption of Maribo occurred at 0.017 MPa and other fragmentations occurred at 0.25 – 4.3 MPa.

The main disruption of Flensburg started at 0.7 MPa and continued until about 2 MPa. The interval of dynamic pressures causing fragmentation was therefore much smaller than that of Maribo. Judging from the light curve, a small fragmentation may have occurred also at the beginning of the bolide (where the brightness increased above the detection limit), i.e. at a height of ~71 km and a dynamic pressure of ~ 0.025 MPa, but only a small amount of mass (~ 1%) may be lost there. We conclude that the Flensburg meteoroid was therefore more homogeneous than Maribo. In this respect it more resembles the Romanian superbolide of January 7, 2015 (Borovička et al. 2017), where, however, no meteorites were recovered and the whole body was probably completely pulverized in the atmosphere.

Additional information about meteoroid fragmentation can be obtained from bolide images. Figure 11 shows selected frames from a casual video taken during a boat trip to the Wangerooge Island, i.e. relatively close to the bolide (~ 120 km). The video could not be used for trajectory determination because it was taken from the sea on an unstable boat with no fixed reference objects. Moreover, the exact position of the boat is unknown. The video shows the bolide before the main flare, at heights of about 54 – 45 km, including the onset of the flare. During the maximum phase, the bolide was hidden behind a boat passenger and emerged again when it was at about 39 km. The heights were determined from the relative time and the time-height relation known from other videos. The video time was related to the other videos using the onset of the main flare.

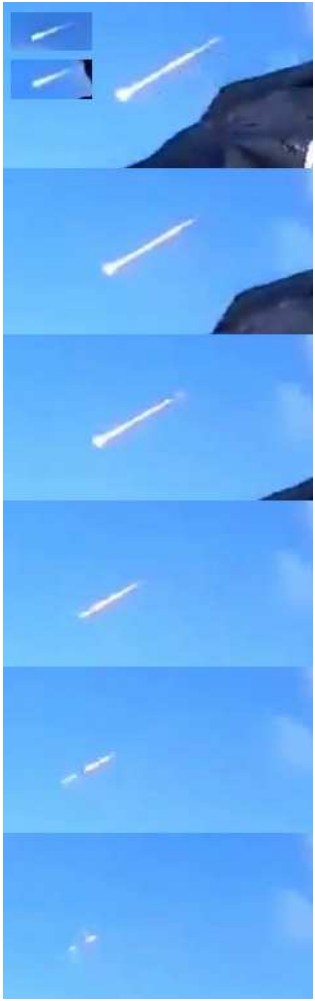

Fig. 11. Bolide images from a video taken at a boat to Wangerooge Island. The six main images show the final phase of the bolide in 0.1 s intervals. The bolide head was at a height of about 38.5 km in the first image (at the top). The dark object is a shoulder of a boat passenger. The two insets show earlier phases of the bolide using the same scale, at heights of about 51 and 47 km, respectively. Author of the video: The Oldenburg Red Cross. The video was downloaded from the site of The Weather Network (www.theweathernetwork.com; also available at https://www.youtube.com/watch?v=30XQhLKOa9k).

As it can be seen in Fig. 11, the bolide was an elongated object before the main flare. It means that some smaller fragments were trailing the main body. The initial fragmentation may have been therefore more severe that it seems from the light curve. The main flare started with brightening of the bolide head, i.e. fragmentation of the main body. When the bolide again emerged on the video after the maximum, it had a bright head and a long and bright wake. After a short time, the head faded and separated from the wake. During the final phase, no forward motion was apparent any more. Instead, two separated sections of the stationary trail faded gradually. This behavior suggests that no large fragments were formed and the meteoroid disintegrated into small (sub kilogram) fragments and dust. However, because of the daylight conditions and thus limited sensitivity of the camera, the presence of an individual large fragment (~ 10 kg, see also the next section) at the end cannot be strictly excluded.

## MASS AND STREWN FIELD

Using the total impact energy of $2 \times 10^{12}$ J reported by the USGS and the known initial velocity of 19.43 km s$^{-1}$, the meteoroid initial mass can be calculated to 10,000 kg. If the light curve in Fig. 10 is fitted by the semi-empirical model in a similar way as Maribo (Borovička et al. 2019), the value of 22,000 kg is obtained from the (generally uncertain) brightness before the main flare. Taking these two values as the possible mass range and assuming bulk density between 1600 – 2000 kg m$^{-3}$ (the meteorite density was 1984 kg m$^{-3}$), the resulting initial equivalent diameter of the meteoroid is between 2 – 3 meters.

The available data are not sufficient to rigorously determine the total fallen mass and/or the number of meteorites. Since the fragmentation was enormous and the bolide terminated at relatively large height of 35 km, the recovered meteorite of 24.5 g may belong to the largest meteorites which reached the ground. Then, taking into account the large initial mass and assuming a regular meteorite mass distribution, hundreds of meteorites of about 10 grams and tens of thousands of about one gram may exist.

Dark flight computation was performed to find the expected spread of the meteorites on the ground. Meteorite fragments were assumed to be formed at two fragmentation points at the heights of 45.5 and 42.5 km, forming the wide bolide maximum, and at a final fragmentation of the remaining piece at the height of 37 km, forming a smaller flare on the light curve (see Fig. 10). The ablation of the formed fragments was computed until their velocities dropped to 2.5 km s$^{-1}$. The ablation coefficient of 0.005 s$^2$/km$^2$ was used, the resulting strewn field is, nevertheless, not sensitive to the exact value of the ablation coefficient. Meteoroid density was assumed to be 2000 kg m$^{-3}$ and the value of ΓA (product of drag coefficient and shape coefficient) was assumed to be 0.8. Other values of the density and ΓA would produce identical strewn field but the meteorite masses at a given location would be different. The high altitude winds and other atmospheric parameters for dark flight computation were taken from two numerical weather prediction models, ECMWF[1] and ALADIN[2]. Both models gave similar results. ECMWF resulted in position 300 – 400 m more to the north than ALADIN. This value is small in comparison with the expected meteorite spread and, moreover, the difference lies in the same direction as the general orientation of the strewn field, so there were no practical consequences for meteorite searches.

The expected strewn field for meteorite masses 5 – 100 g is shown in Fig. 12. The strewn field is oriented from the south to the north. Larger meteorites lie more to the north, a consequence of the direction of bolide flight. The direction of the strongest winds was from WNW, so the whole strewn field was shifted to the east from the prolonged bolide trajectory. Meteorites originating at the lower height of 37 km are shifted to the north (more exactly, in the azimuth of bolide flight) in respect to the meteorites of the same mass originating higher, simply because they started the dark flight later. As a consequence, the strewn field of meteorites originating at 37 km lies more to the east as of those originating higher, as shown in Fig. 12. When moving from west to east, one should first encounter larger meteorites originating at higher heights (if they exist) and then smaller meteorites from lower heights.

The line for meteorites originating at lower heights passes very close to the position of the recovered meteorite. Although meteorites of masses of about 45 g were predicted there and the actual meteorite had only 24.5 g, that discrepancy can be easily removed by adjusting ΓA to 0.65 instead of 0.8 for this meteorite, which is quite plausible. For a perfect sphere, the value of ΓA would be ~0.6 (A=1.21, Γ~0.5). We can therefore conclude with high degree of reliability that the meteorite originated from a part of the meteoroid, which survived the main disruption at heights above 40 km and fragmented at lower heights around 37 km at a dynamic pressure about 2 MPa.

---

[1] European Centre for Medium-Range Weather Forecasts
[2] Aire Limitée Adaptation dynamique Développement InterNational

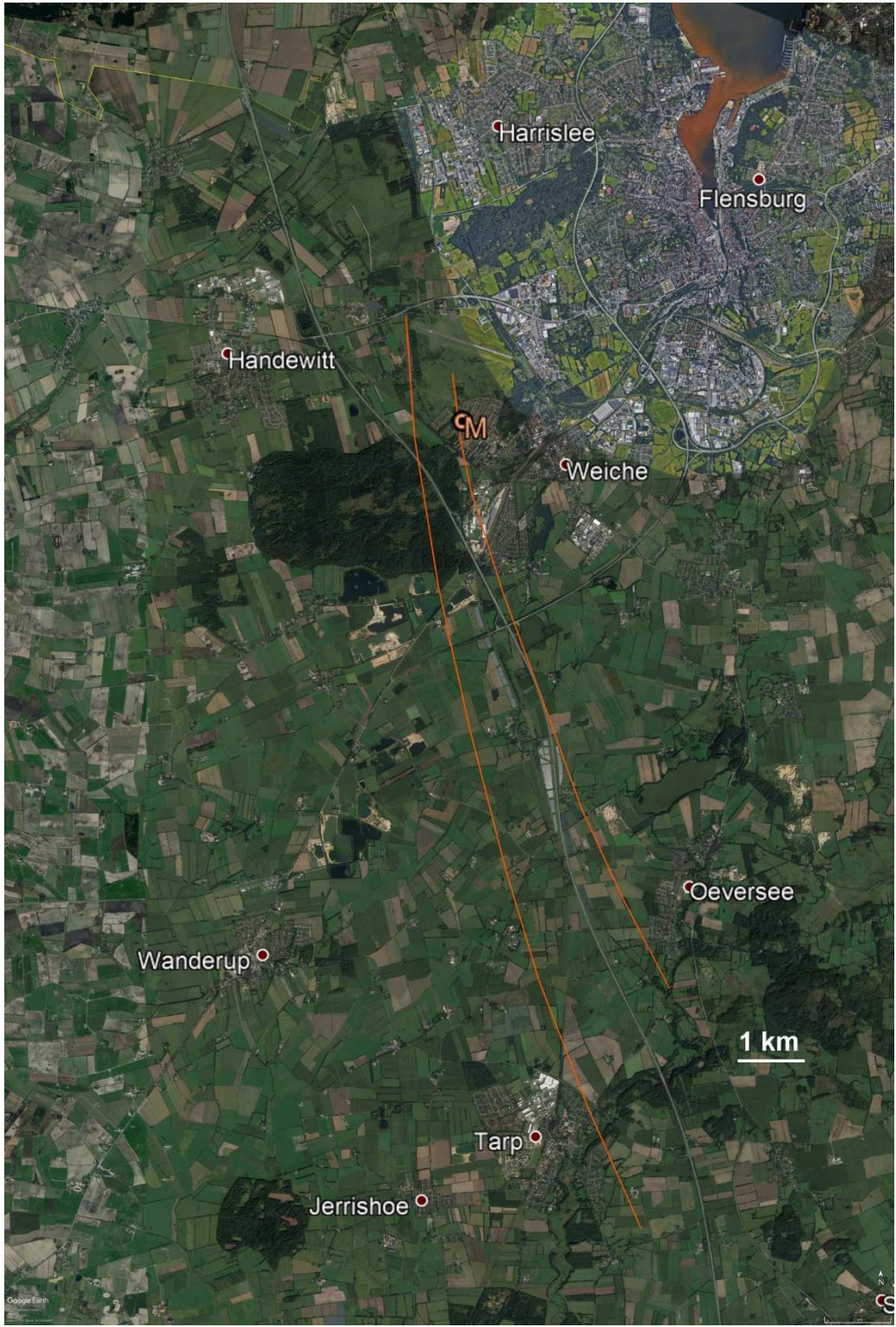

Fig. 12. Map of the expected meteorite fall area near Flensburg. North is up. The position of the only recovered meteorite is marked M. The modeled meteorite position are depicted by two orange lines. Meteorites originating in the main fragmentation event at the heights of 43 – 46 km should be spread along the left (western) line. Meteorite masses increase from the south to the north. The extent corresponds to meteorite masses from about 5 grams to 100 grams, assuming a meteorite density of 2000 kg m$^{-3}$ and ΓA=0.8. Meteorites of masses 5 – 50 g originating in the last major fragmentation event at a height of ~37 km are expected to be spread along the right line. The source of the background image is Google Earth.

Rather extensive searches for more meteorite fragments remained unsuccessful. First sporadic searches around the position of the first meteorite started ten days after the find. More systematic searches were performed in the second half of November 2019 and then in February 2020 after the bolide trajectory was computed from the first two calibrated videos.

**DISCUSSION**

Although the available bolide records were not ideal, the obtained trajectory and velocity are reliable and the precision is reasonably good. In addition of clarifying the origin of the meteorite, this result also represents one of the few opportunities to compare the velocity vector reported by the USGS with independent data. Any details how the USGS data are obtained have not been published. As the USGS dataset is being used to infer properties of meter-scaled meteoroids (Brown et al. 2016, Wiegert et al. 2020), it is important to know the reliability of the data. In this particular case, the direction of flight (the radiant) was determined well (within one degree). The velocity was by 0.9 km s$^{-1}$ lower than the actual entry velocity. The height of the maximum brightness point was determined well. The geographic position was shifted by 0.3° in longitude (~ 20 km). Overall, the Flensburg belongs to the best observed bolides by the USGS (cf. Devillepoix et al. 2019). Still, if USGS data were used to compute the heliocentric orbit, the resulting eccentricity (0.64), and thus also semimajor axis (2.37 AU) and aphelion distance (3.9 AU) would be too small in comparison with the results derived in this work. The perihelion distance and all angular elements would be nearly identical.

The most important fact revealed by this study is that the Flensburg meteoroid was in the 5:2 resonance with Jupiter located at 2.82 AU. Since the resonance is able to quickly raise the orbital eccentricity and thus bring the meteoroid into an Earth crossing orbit, it explains the extremely short cosmic ray exposure age (CRE) of about 7 kyr (Bischoff et al. 2020). The parent asteroid of Flensburg was very probably orbiting close to the inner or outer edge of the resonance, so that Flensburg entered the resonance soon after being ejected by a collisional event.

Granvik and Brown (2018) evaluated the most likely source regions of 25 meteorites with known orbits. From them, only the H chondrites Košice and Ejby (for more details about their falls see Borovička at al. 2013 and Spurný et al. 2017, respectively) had more than 10% probability of coming through the 5:2 resonance (~18% and 25%, respectively). Košice, however, had a higher probability of coming from the 3:1 resonance located closer to the Sun (at 2.5 AU). For Ejby, the highest probability (~ 55%) was obtained for coming from a Jupiter Family Comet. Since that origin can be effectively excluded for an ordinary chondrite, 5:2 resonance is the most likely source origin of Ejby, although the 3:1 resonance with computed probability of 10-15% cannot be fully excluded. Apart Ejby and Košice, the CM2 chondrites Maribo and Sutter's Mill were the only two meteorites with non-negligible probability of cometary origin on the basis of their orbits. But the 3:1 and ν6 resonances were nearly equally possible in both cases, while the 5:2 resonance was unlikely (~5%). Flensburg is therefore the only meteorite to date clearly coming through the 5:2 resonance. In contrast to its young CRE, Flensburg contains very old carbonates, not affected since their formation 4564 Myr ago (Bischoff et al. 2020). This fact may be related to the location of its parent asteroid farther from the Sun than for most other meteorites.

A strange aspect is the recovery of only a single meteorite. With the initial mass of 10,000 – 20,000 kg, Flensburg was comparable to the carbonaceous chondrites Tagish Lake (50,000-90,000 kg, Brown et al. 2002, Hildebrand et al. 2006) and Sutter's Mill (20,000-80,000 kg, Jenniskens et al. 2012), which both yielded much more meteorites. One possible explanation can be that the recovered meteorite represents an exceptionally compact part of the original meteoroid and the number of fallen meteorites of substantial mass (>1 g) was indeed low. The fact that vast majority of radiation was released above the height of 39 km suggests that only ~3% of the original mass reached the height of 37 km, where the meteorite originated, in form of one or more large bodies (>> 10 kg). The fragmentations at higher altitudes may have produced only dust or meteorites of negligible mass. Most of the small fragments formed at 37 km may have crumbled repeatedly, even during the dark flight, so only few of them reached ground. Unfortunately, the available data are not sufficient to confirm this scenario. In terms of maximum dynamic pressure, the 2 MPa reached by Flensburg is somewhat lower than for Maribo (5 MPa, Borovička et al. 2019) but similar to Tagish Lake (Brown et al. 2002, Popova et al. 2011). Flensburg was therefore not exceptional in its fragmentation strength. Perhaps the more important question is what the outcome of fragmentation is. The Romanian superbolide reached up to 3 MPa and produced no meteorites (Borovička et al. 2017).

Another possibility is that Flensburg produced many meteorites but they degraded quickly in wet European climate. The first meteorite was found the next day after the fall but thorough systematic searches were delayed by two months. Note that Tagish Lake meteorites were preserved in ice (Hildebrand et al. 2006) and other carbonaceous chondrites were typically found either soon or in drier environments.

## CONCLUSIONS

Flensburg is a unique carbonaceous meteorite. It was therefore important to determine its pre-fall heliocentric orbit. The results showed clearly that the meteorite originated in the vicinity of the 5:2 resonance with Jupiter at a heliocentric distance of 2.82 AU. It is in agreement with its very short CRE, which demonstrates that no slow drift mechanism such as the Yarkovsky effect was involved in the transport of the meteorite to the Earth. Flensburg seems to originate from a larger heliocentric distance than other meteorites with known orbit. The recovery of a single meteorite suggests the ratio of fallen to initial mass was quite low even in comparison with other carbonaceous chondrites. The recovery was enabled by the combination of a large initial mass and the fall into an inhabited area. This fortunate coincidence may explain why no other meteorite of this kind is known.

*Acknowledgements*. –We thank the authors of the casual videos for making them available on the Internet. The mindfulness of Erik Due-Hansen and his exemplary cooperation with the scientists is highly appreciated. R. Brožková from the Czech Hydrometeorological Institute is acknowledged for providing the data from the ALADIN weather model. This work was supported by grant no. 19-26232X from the Czech Science Foundation.

## REFERENCES

Bischoff A., Alexander C., Barrat J.-A., Burkhardt C., Busemann H., Degering D., Di Rocco T., Fischer M., Fockenberg T., Foustoukos D. I., Gattacceca J., Godinho J. R. A., Harries D., Heinlein D., Hellmann J. L., Hertkorn N., Holm A., Jull A. J. T., Kerraouch I., King A. J., Kleine T., Koll

D., Lachner J., Ludwig T., Merchel S., Mertens C., Morino P., Neumann W., Pack A., Patzek M., Pavetich S., Reitze M. P., Rüfenacht M., Rugel G., Schmidt C., Schmitt-Kopplin P., Schönbächler M., Trieloff M., Wallner A., Wimmer K., and Wölfer E. 2020. The old, unique C1 chondrite Flensburg – insight into the first processes of aqueous alteration, brecciation, and the diversity of water-bearing parent bodies and lithologies. *Geochim. Cosmochim. Acta* – Submitted

Borovička J. 1990. The comparison of two methods of determining meteor trajectories from photographs. *Bulletin of the Astronomical Institutes of Czechoslovakia* 41:391-396.

Borovička J., Tóth J., Igaz A., Spurný P., Kalenda P., Haloda J., Svoreň J., Kornoš L., Silber E., Brown P., and Husárik M. 2013. The Košice meteorite fall: Atmospheric trajectory, fragmentation, and orbit. *Meteoritics and Planetary Science* 48:1757-1779.

Borovička J. 2014. The analysis of casual video records of fireballs. In *Proceedings of the International Meteor Conference 2013, Poznań, Poland*, edited by Gyssens, M. Roggemans, P., and Żołądek P., pp. 101-105.

Borovička J., Spurný P., Grigore V. I., and Svoreň J. 2017. The January 7, 2015, superbolide over Romania and structural diversity of meter-sized asteroids. *Planetary and Space Science* 143: 147-158.

Borovička J., Popova O., and Spurný P. 2019. The Maribo CM2 meteorite fall—Survival of weak material at high entry speed. *Meteoritics and Planetary Science* 54:1024-1041.

Brown P. G., Hildebrand A. R., Zolensky M. E., Grady M., Clayton R. N., Mayeda T. K., Tagliaferri E., Spalding R., MacRae N. D., Hoffman E. L., Mittlefehldt D. W., Wacker J. F., Bird J. A., Campbell M. D., Carpenter R., Gingerich H., Glatiotis M., Greiner E., Mazur M. J., McCausland P. J., Plotkin H., and Rubak Mazur T. 2000. The Fall, Recovery, Orbit, and Composition of the Tagish Lake Meteorite: A New Type of Carbonaceous Chondrite. *Science* 290:320-325.

Brown P. G., ReVelle D. O., Tagliaferri E., and Hildebrand A. R. 2002. An entry model for the Tagish Lake fireball using seismic, satellite and infrasound records. *Meteoritics and Planetary Science* 37:661-675.

Brown P., Wiegert P., Clark D., and Tagliaferri E. 2016. Orbital and physical characteristics of meter-scale impactors from airburst observations. *Icarus* 266:96-111.

Ceplecha Z. 1987. Geometric, dynamic, orbital and photometric data on meteoroids from photographic fireball networks. *Bulletin of the Astronomical Institutes of Czechoslovakia* 38: 222-234

Devillepoix H. A. R., Bland P. A., Sansom E. K., Towner M. C., Cupák M., Howie R. M., Hartig B. A. D., Jansen-Sturgeon T., and Cox M. A. 2019. Observation of metre-scale impactors by the Desert Fireball Network. *Monthly Notices of the Royal Astronomical Society* 483:5166-5178.

Granvik M., and Brown P. 2018. Identification of meteorite source regions in the Solar System. *Icarus* 311: 271-287.

Hankey M., Perlerin V., and Meisel D. 2020. The all-sky-6 and the Video Meteor Archive system of the AMS Ltd.. *Planetary and Space Science* 190:105005.

Hildebrand A. R., McCausland P. J. A., Brown P. G., Longstaffe F. J., Russell S. D. J., Tagliaferri E., Wacker J. F., Mazur M. J. (2006). The fall and recovery of the Tagish Lake meteorite. *Meteoritics and Planetary Science* 41: 407–431.

Jenniskens P., Fries M. D., Yin Q.-Z., Zolensky M., Krot A. N., Sandford S. A., Sears D., Beauford R., Ebel D. S., Friedrich J. M., Nagashima K., Wimpenny J., Yamakawa A., Nishiizumi K., Hamajima Y., Caffee M. W., Welten K. C., Laubenstein M., Davis A. M., Simon S. B., Heck P. R., Young E. D., Kohl I. E., Thiemens M. H., Nunn M. H., Mikouchi T., Hagiya K., Ohsumi K., Cahill T. A., Lawton J. A., Barnes D., Steele A., Rochette P., Verosub K. L., Gattacceca J., Cooper G., Glavin D. P., Burton A. S., Dworkin J. P., Elsila J. E., Pizzarello S., Ogliore R., Schmitt-Kopplin P., Harir M., Hertkorn N., Verchovsky A., Grady M., Nagao K., Okazaki R., Takechi H., Hiroi T., Smith K., Silber E. A., Brown P. G., Albers J., Klotz D., Hankey M., Matson R., Fries J. A., Walker R. J., Puchtel I., Lee C.-T. A., Erdman M. E., Eppich G. R., Roeske S., Gabelica Z., Lerche M., Nuevo M., Girten B., and Worden S. P. 2012. Radar-Enabled Recovery of the Sutter's Mill Meteorite, a Carbonaceous Chondrite Regolith Breccia. *Science* 338:1583.

Moons M. 1996. Review of the dynamics in the Kirkwood gaps. *Celestial Mechanics and Dynamical Astronomy* 65: 175–204.

Popova O., Borovička J., Hartmann W. K., Spurný P., Gnos E., Nemtchinov I., and Trigo-Rodríguez J. M. 2011. Very low strengths of interplanetary meteoroids and small asteroids. *Meteoritics and Planetary Science* 46:1525-1550.

Spurný P., Borovička J., Baumgarten G., Haack H., Heinlein D., and Sørensen A. N. 2017. Atmospheric trajectory and heliocentric orbit of the Ejby meteorite fall in Denmark on February 6, 2016. *Planetary and Space Science* 143:192-198.

Tancredi G. 2014. A criterion to classify asteroids and comets based on the orbital parameters. *Icarus* 234: 66-80.

Wiegert P., Brown P., Pokorný P., Ye Q., Gregg C., Lenartowicz K., Krzeminski Z., and Clark D. 2020. Supercatastrophic Disruption of Asteroids in the Context of SOHO Comet, Fireball, and Meteor Observations. *Astronomical Journal* 159:143.